%% file: main.tex
\documentclass[10pt,conference]{IEEEtran}

\usepackage{amsmath,amsfonts}
\usepackage{graphicx}
\usepackage{textcomp}
\usepackage{cite}
\usepackage{multirow}
\usepackage{tabularx}
\usepackage{booktabs}
\usepackage{amsmath}
\usepackage{amsthm}
\usepackage{color}
\usepackage{enumitem}
\usepackage{caption}
\usepackage{subcaption}
\usepackage{listings}
\usepackage{url}
\usepackage{pgfplots}
\usepackage{color}
\usepackage{xspace}

\def\BibTeX{{\rm B\kern-.05em{\sc i\kern-.025em b}\kern-.08emT\kern-.1667em\lower.7ex\hbox{E}\kern-.125emX}}
\DeclareCaptionFont{black}{\color{black}}
\DeclareCaptionFormat{listing}{\colorbox{gray!50}{\parbox[l]{\columnwidth}{#1#2#3}}}
\begin{document}

%
\newcommand{\toolname}{\textsc{Hawkeye}\xspace}
\title{Hawkeye: Change-targeted Testing for Android Apps based on Deep Reinforcement Learning}

%
\author{\IEEEauthorblockN{Chao Peng$^1$,
Zhengwei Lv$^1$,
Jiarong Fu$^1$,
Jiayuan Liang$^1$,
Zhao Zhang$^1$,
Ajitha Rajan$^2$,
Ping Yang$^1$
}
\IEEEauthorblockA{
$^1$\textit{ByteDance, Beijing, China}\\
$^2$\textit{University of Edinburgh, Edinburgh, United Kingdom}\\
Email:~\{{\{pengchao.x, lvzhengwei.m, fujiarong, liangjiayuan.522, zhangzhao.a, yangping.cser\}@bytedance.com}\\arajan@ed.ac.uk}
}


\maketitle

\begin{abstract}
\input{abstract}
\end{abstract}

\begin{IEEEkeywords}
    software testing, deep reinforcement learning, android, graphical user interface
\end{IEEEkeywords}

\section{Introduction}
\label{sec:introduction}
\input{introduction}

\section{Background}
\label{sec:background}
\input{background}

\section{Related Work}
\label{sec:related}
\input{related}

\section{Our Approach}
\label{sec:approach}
\input{approach}

\section{Experiment}
\label{sec:experiment}
\input{experiment}

\section{Results}
\label{sec:result}
\input{result}

\section{Conclusion}
\label{sec:conclusion}
\input{conclusion}

\bibliographystyle{IEEEtran}
\bibliography{reference}

\end{document}

%% file: abstract.tex
Android Apps are frequently updated to keep up with changing user, hardware, and business
demands.
Ensuring the correctness of App updates through extensive testing is crucial to avoid potential bugs reaching the end user. 
Existing Android testing tools generate GUI events focussing on improving the test coverage of the entire App rather than prioritising updates and its impacted elements.
Recent research has proposed change-focused testing but relies on random exploration to exercise the updates and impacted GUI elements that is ineffective and slow for large complex Apps with a huge input exploration space. 
We propose directed testing of App updates with \toolname that is able to prioritise executing GUI actions associated with code changes based on deep reinforcement learning from historical exploration data.
Our empirical evaluation compares \toolname with state-of-the-art model-based and reinforcement learning-based testing tools FastBot2 and ARES using 10 popular open-source and 1 commercial App. We find that \toolname is able to generate GUI event sequences targeting changed functions more reliably than FastBot2 and ARES for the open source Apps and the large commercial App. \toolname achieves comparable performance on smaller open source Apps with a more tractable exploration space.
The industrial deployment of \toolname in the development pipeline also shows that \toolname is ideal to perform smoke testing for merge requests of a complicated commercial App.

%% file: introduction.tex
The rapid growth of mobile technology has led to an unprecedented increase in the development and usage of mobile Apps, particularly in the Android ecosystem.
These Apps are frequently updated (typically weekly) to keep up with changing user, hardware and business demands. 
To ensure security and correctness, updates in Apps need to be tested thoroughly to ensure changes and 
existing functionality work as expected.

Traditional testing of Android Apps relies on human interaction with the Apps to identify unexpected behaviours which is a time-consuming and error-prone process.
Several automated mobile App testing techniques have been proposed in the
literature~\cite{choudhary2015automated,borges2018droidmate, su2017guided,baek2016automated,takala2011experiences, su2016fsmdroid, riganelli2020data, choi2018detreduce, mirzaei2016reducing, song2017ehbdroid}. Majority of existing work focuses on testing only one version
of a mobile App and are not applicable to App updates. 
There is work on regression test selection for App updates that chooses a subset of tests exercising the updates from an existing test suite~\cite{sharma2019qadroid, jiang2018retestdroid, do2016redroid, choi2018detreduce, li2017atom, do2016regression}. QADdroid~\cite{sharma2019qadroid} is the only tool in literature that considers changes and their impact at the GUI level when selecting regression tests. Regression test selection techniques, however, only select tests from an existing test suite; they do not generate new tests that exercise changes. To the best of our knowledge, CAT~\cite{peng2021cat} is the only tool in the literature that generates GUI events to exercise changes. However, it relies on random GUI exploration before reaching the change-impacted GUI element. If the App under test is complex, the random execution can take hours, and as complexity increases, it may take an impractical amount of time before the target GUI element is reached, rendering it ineffective in real-world practice. 

\begin{figure}
    \centering
    \includegraphics*[scale=0.55]{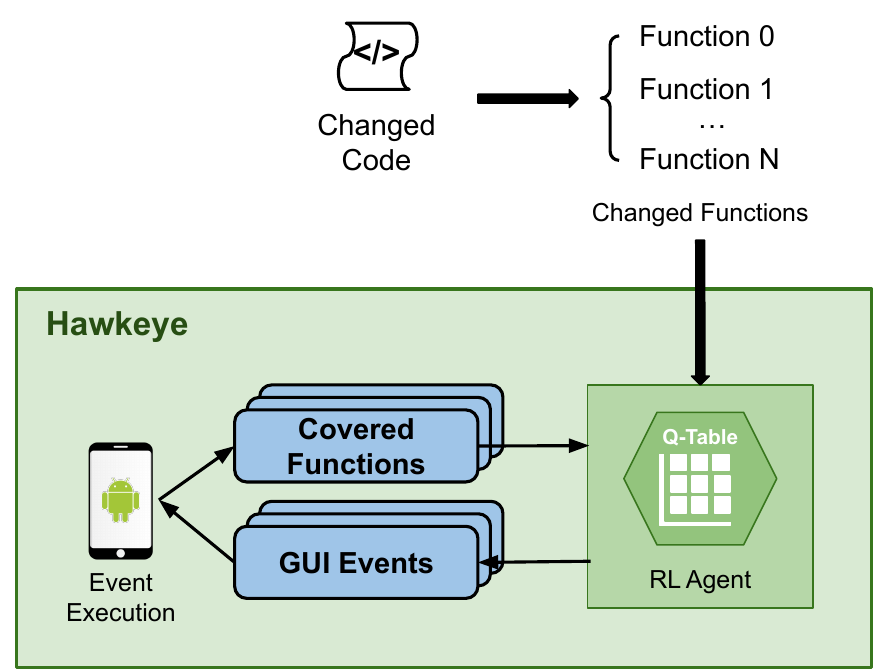}
    \caption{Overview of \toolname}
    \label{fig:introduction}
\end{figure}

In this paper, we present \toolname, a novel approach focusing on generating GUI events to quickly exercise App changes based on reinforcement learning. The overview of \toolname is shown in Figure~\ref{fig:introduction}. \toolname is triggered when a merge request is submitted. It first analyses the codebase and the code commit to identify changed functions and initiates a GUI exploration task to perform change-targeted testing.
To quickly locate and exercise changed functions, a reinforcement learning model is loaded from the historical GUI exploration data to infer GUI events that are most likely to executed these functions.
The task is terminated once all changed functions are covered or the execution times out. 
Additionally, \toolname uses a client server implementation with the ability to test the same App on multiple devices simultaneously while sharing the deep reinforcement learning agent.

We assess the utility and efficacy of \toolname for testing App updates using a dataset consisting of 10 open-source Android Apps from the F-Droid App market, along with a commercial enterprise collaboration App, Lark, developed by an industrial partner. We benchmark \toolname's performance against two reinforcement learning-based GUI testing tools for Android, namely \textit{ARES}~\cite{romdhana2022deep} and \textit{Fastbot2}~\cite{lv2022fastbot2}. 

Experimental results show that \toolname can exercise changed functions more frequently and reliably than \textit{Fastbot2} and \textit{ARES} using fewer GUI events. We also find that for testing changes in the complex commercial App, \toolname achieves better performance. 
85\% of the changed functions can be covered within the first 5 minutes of the 180-minute testing run. We find that \toolname finds it easier to exercise functions that were seen more frequently during model training.

In summary, the main contributions in this paper are:
\begin{enumerate}
    \item A novel deep reinforcement learning-based GUI testing tool that prioritises GUI events related to changed functions.
    \item An optimised client-server implementation that allows testing on multiple devices while sharing the deep reinforcement learning agent.
    \item Empirical evaluation against state-of-the-art reinforcement learning tools using ten open-source and one commercial App.
\end{enumerate}

%% file: background.tex
Before we discuss our approach in detail, we briefly introduce basic concepts in Android App development and testing. We also provide background in reinforcement learning and how it can be used for test generation. 

\subsection{Android Apps}
\label{sec:android-platform}

Android Apps are typically developed using Java or Kotlin, which are compiled into Java bytecode and then converted to Dalvik bytecode. To enhance performance, native code can also be incorporated. Android packaging tools such as Gradle build the Dalvik bytecode, native code (if available), and any data or resource files into an Android package (APK), which is an archive file with an .apk extension. The APK file is the sole requirement for installing the App on Android devices.

In order to construct the APK file, an Android project relies on these components: 
\begin{enumerate}
    \item [(i)] source code files that contain the code for the App's classes and functions,
    \item [(ii)] layout-XML files that outline the graphical user interface layout for all activities, and
    \item [(iii)] the Android manifest, which is located in the App's root folder as AndroidManifest.xml and provides crucial details about the App, such as the package name (used to find the source code), component lists, required user permissions, utilized hardware and software features, and necessary API libraries.
\end{enumerate}
\subsection{Android GUI and Testing}
\label{sec:terminology}

The window displayed in the screen is referred to as an \texttt{Activity}, encompassing a range of \texttt{GUI elements} (also known as \texttt{Views} or \texttt{Widgets}) such as buttons and text fields.
By incorporating suitable callbacks for each life-cycle stage (i.e., created, paused, resumed, and destroyed), developers can manage the behavior of individual Activities. These Activities must be initially specified in the \textit{AndroidManifest.xml} file and are executed as Java classes within the source code directory.

\texttt{GUI elements} serve as the fundamental building blocks for user interaction, such as textboxes, buttons, and containers for other \texttt{GUI elements}. 
\texttt{GUI elements} play a crucial role in event handling, which may include button clicks, text editing, touch events, and more. To react to a specific event type, it needs to register a suitable event listener and implement the associated callback method (invoked by the Android Framework when the \texttt{GUI element} is activated through user interaction). For instance, to enable a button to respond to a click event, it must register the View.onClickListener event listener and implement the relevant onClick() callback method. Upon detecting a button click event, the Android framework will call the onClick() method for that specific \texttt{GUI element}.

An \texttt{event sequence} refers to an organized series of input events. In this paper, the term \texttt{state} denotes the GUI state, which encompasses the current screen's GUI information and all its elements. Since Android Apps are event-driven, inputs typically take the form of events. Manually writing or recording input events can be laborious and time-intensive~\cite{sharma2014quantitative}. Consequently, automated input event generation for testing Android Apps has become a thriving research area. The following section provides an overview of existing research in Android testing.

\subsection{Reinforcement Learning}

Reinforcement Learning (RL) is a sub-domain of machine learning that focuses on training agents to make intelligent decisions by interacting with an environment.
In RL, an agent learns to select actions within a given context to maximize a cumulative reward signal.
The underlying idea is that the agent learns an optimal policy through trial-and-error, wherein it continuously updates its knowledge based on the observations and feedback obtained from the environment.
The agent's learning process is driven by balancing the exploration of the environment and the exploitation of the acquired knowledge.
Reinforcement learning has exhibited impressive results in a wide variety of domains, ranging from robotics and game-playing to recommendation systems and self-driving cars.
In the context of automated Android App testing, reinforcement learning can be leveraged as a means to efficiently explore the state-space of an application by formulating it as a decision-making problem.

%% file: related.tex
In this Section, we provide an overview of current research on Android GUI test generation, which is divided into random, model-driven and learning-based testing approaches. Additionally, we discuss related studies that focus on choosing regression tests or generating new tests based on App updates.. 


\subsection{Random Android GUI Testing} Android Monkey~\cite{androidmonkey} is a well-known and state-of-the-practice random testing tool that analyses the GUI and arbitrarily selects events to be executed in the current state until the number of executed events surpasses the user-defined limit. DynoDroid~\cite{machiry2013dynodroid} employs heuristics for input event selection instead of complete randomness. However, DynoDroid has not been updated for several years and only supports Android version 2.3.5 (with Android 13 being the latest version). Wetzlmaier et al.~\cite{wetzlmaier2017hybrid} enhance existing test inputs by incorporating random test inputs, providing users with more control than Monkey. None of the current random testing tools concentrate on App updates.

\subsection{Model-based Android Testing} DroidBot~\cite{li2017droidbot} and DroidMate~\cite{jamrozik2016droidmate, borges2018droidmate} emphasise generating test inputs based on GUI models. DroidMate directs test input generation in real-time using the GUI model. DroidBot consults the GUI model of the target App, calculates, and performs possible events within this model. DroidBot also offers a user-friendly interface for App exploration.

In contrast to static GUI model-based test generation, Ape~\cite{gu2019practical} dynamically optimises the GUI model by taking advantage of runtime information during testing. Ape employs a decision tree-based representation while exploring the App and continuously refines the GUI model, aiming to maintain an optimal balance between model size and model accuracy.

\subsection{Machine Learning-based Android Testing} Machine learning techniques have been extensively applied in testing Android applications. Some methods~\cite{borges2018guiding, li2019humanoidupdated} use supervised learning to generate test inputs that achieve high coverage. Borges et al.~\cite{borges2018guiding} suggest a straightforward yet effective approach that can guide test generation towards UI elements with the highest likelihood of being reactive. Humanoid~\cite{li2019humanoidupdated} utilizes a deep neural network to learn about users' interactions with the Apps being tested, enabling it to prioritise user-preferred inputs for new UIs

In addition, many researchers have tried to adopt reinforcement learning in automatic test input generation.
Q-testing~\cite{pan2020reinforcement} is a Q-learning based approach for automated testing. Q-testing leverages Q-table as a lightweight model while exploring unfamiliar functionalities with
a curiosity-driven strategy.
ARES~\cite{romdhana2022deep} and ATAC~\cite{gao2023deep} are both testing tools that utilise deep reinforcement learning algorithms. ARES employs algorithms such as DDPG, SAC and TD3 as agents to learning both the state similarity and action-value functions, while ATAC leverages the Advantage Actor-Critic algorithm to generate test cases for Android GUI testing.
DeepGUI~\cite{collins2021deep} is a Deep Q-Network-based testing tool. It can generate diverse test cases by exploring the application under test using trial-and-error, with the aim of achieving code coverage and revealing failures and crashes.
All three tools --- ARES, ATAC and DeepGUI --- do not consider changes in Android Apps which is the focus of our approach. We compare performance against ARES in Section~\ref{sec:result}. We were unable to evaluate against ATAC as it is not publicly available. We attempted to use DeepGUI in our experiments but the test generation was extremely time consuming making it impractical to use.

\subsection{Regression Test Selection and Generation} 
\label{sec:regression}

Numerous studies have investigated the selection of regression tests based on App updates and their effects. However, \toolname's focus is different - it aims at \emph{generating inputs} to exercise changed functions unlike existing regression test selection contributions that focus on test selection. 

Redroid~\cite{do2016redroid, do2016regression} and ReTestDroid~\cite{jiang2018retestdroid} are regression test selection techniques that compare Java source files from the original and updated App versions to identify changes and calculate change impact at the source code level. The tools select regression tests that exercise change-impacted code. ReTestDroid handles more Java features than Redroid, such as fragments, native code, and asynchronous tasks. Both tools perform change impact analysis at the source code level, not considering GUI elements, and are used for test selection rather than generation. 

QADroid~\cite{sharma2019qadroid} and ATOM~\cite{li2017atom} also carry out test selection for regression versions of Apps. QADroid analyses the impact of App updates on code and GUI elements. QADroid constructs call graphs based on FlowDroid~\cite{arzt2014flowdroid} and links events to function calls using event-function bindings defined in the source code. QADroid does not support change impact analysis for dynamic GUI elements, as it does not support Java reflection. ATOM~\cite{li2017atom} creates an event-flow graph for each App version, with nodes representing activities and edges representing events that trigger activity transitions. It then computes a delta graph using event-flow graphs of the updated and original App versions. Only events present in the delta graph are chosen for regression test selection. 

CAT~\cite{peng2021cat} conducts change impact analysis at both the source code and GUI levels, and prioritises change-related GUI events. However, CAT does not provide a way to localise these GUI events and relies on purely random GUI exploration before these events are available on the screen. More importantly, the static analysis on the source code is not applicable to large Apps. CAT is not applicable to large industrial Apps that we used in our initial experiment on a devbox with 128 GB memory. As a result, we do not use this tool in our evaluation.  

ATUA~\cite{ngo2022atua} is a state-of-the-art update-driven App testing tool that achieves high coverage of the updated code with a minimal number of test inputs. It employs a model-based approach that combines static and dynamic program analysis to select test inputs that exercise the updated methods.
Similar to CAT, the static analysis phase employed by ATUA to perform change impact analysis and widget transition graph construction fail on Apps larger than 20 MB and we have to exclude this tool in our evaluation.

%% file: approach.tex
\begin{figure*}[h]
    \centering
    \includegraphics[scale=0.45]{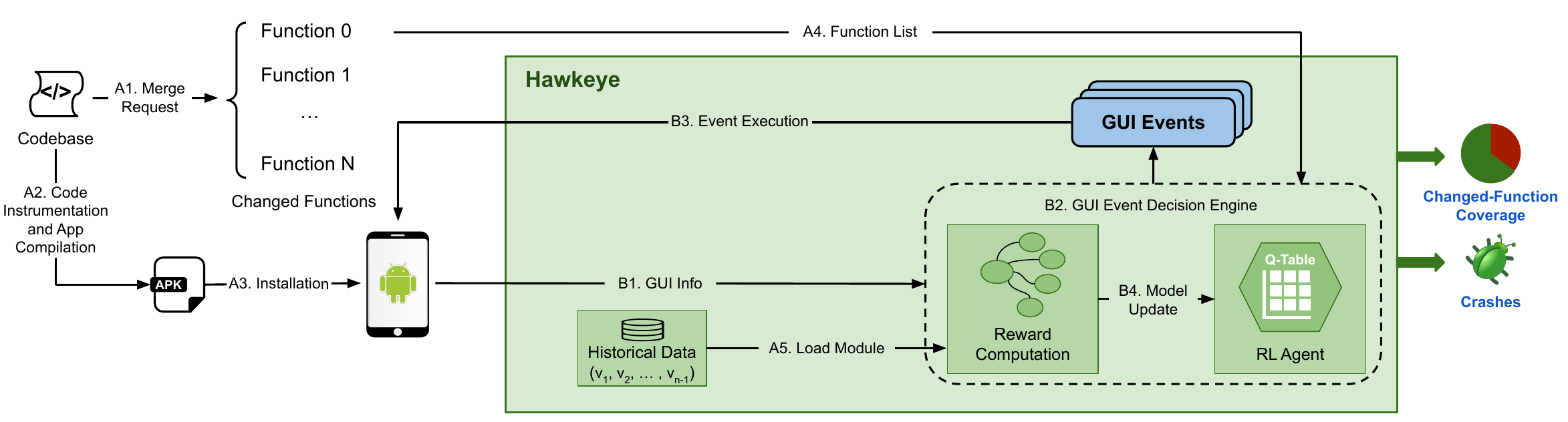}
    \caption{\toolname Workflow}
    \label{fig:workflow}
\end{figure*}

In this section, we present \toolname based on deep reinforcement learning model to guide change-targeted GUI exploration. The workflow of \toolname is illustrated in Figure~\ref{fig:workflow}. The input to \toolname is the list of changed functions and the instrumented App with real-time function coverage monitoring and the output is GUI event sequences to be executed. 

\toolname includes two major phases for each testing
run. The first phase (with Prefix A in Figure~\ref{fig:workflow}) does setup \emph{before} testing:
\begin{enumerate}
    \item[A1.] The list of changed functions (target functions) is generated by analyzing the merge request from the codebase. The user can also provide this list of target functions manually in the tool configuration file.
    \item[A2.] \toolname instruments the App on the source code level to enable the compiled APK file capable of recording and reporting real-time function coverage. The coverage information is used to train the reinforcement learning model and determine whether any of target functions is covered.
    \item[A3.] The instrumented App is installed on multiple devices to perform GUI exploration. During testing, the device continuously sends GUI info to \toolname and execute selected GUI events chosen by \toolname.
    \item[A4.] The list of target functions is sent to the decision engine to infer GUI events to be executed.
\end{enumerate}

The second phase (with prefix B) performs guided GUI exploration.
\begin{enumerate}
    \item[B1.] \toolname dumps the current GUI screen from the App under test (AUT).  \toolname communicates with the device via the Android Debug Bridge to query GUI info and gather executable GUI events available on the current screen.
    \item[B2.] \toolname selects a GUI event with the most likelihood to exercise one of the target functions.
    \item[B3.] The selected GUI event is executed on the device and the GUI state change is sent to the decision engine to update models (B4).
\end{enumerate}

These steps (“B1”-“B4”) is executed in a loop until the threshold of number of GUI events is reached or the time budget is used up.

In the rest of this section, we discuss each step in detail.

\subsection{Input Preprocessing}
\label{sec:preprocessing}

Mapping between GUI actions and source code functions is essential to train the deep reinforcement learning model. We obtain this mapping by instrumenting the Apps to monitor functions executed. 
We build the instrumentation tool as a Gradle plugin based on ASM\footnote{\url{https://asm.ow2.io/}}, a Java bytecode manipulation and analysis framework.
Gradle is the default build automation tool used by the official Android development environment, Android Studio.
As a gradle plugin, code instrumentation can be integrated into existing Android projects seamlessly by only adding the name of the tool to the plugin list of the project configuration file.

\begin{lstlisting}[caption={Example Instrumented Methods},label={lst:instrumentation}]
    void foo() {
        // function ID = 1;
        Hawkeye.logMethod(1);
        ...
    }

    void bar() {
        // function ID = 2;
        Hawkeye.logMethod(2);
        ...
    }
\end{lstlisting}

As shown in Listing~\ref{lst:instrumentation}, the code instrumentation inserts a function call to \texttt{logMethod} to the beginning of all methods. The implementation of this function is sending a socket message to \toolname with the method ID.
After code instrumentation, the mapping between function signature and function ID is stored in the database for \toolname to query.
As a result, during testing, after each GUI event is exercised, \toolname is aware of the functions that get triggered.

\subsection{RL-based Guided GUI Exploration}
\label{sec:rl}

\begin{figure*}[htb]
    \centering
    \includegraphics[scale=0.55]{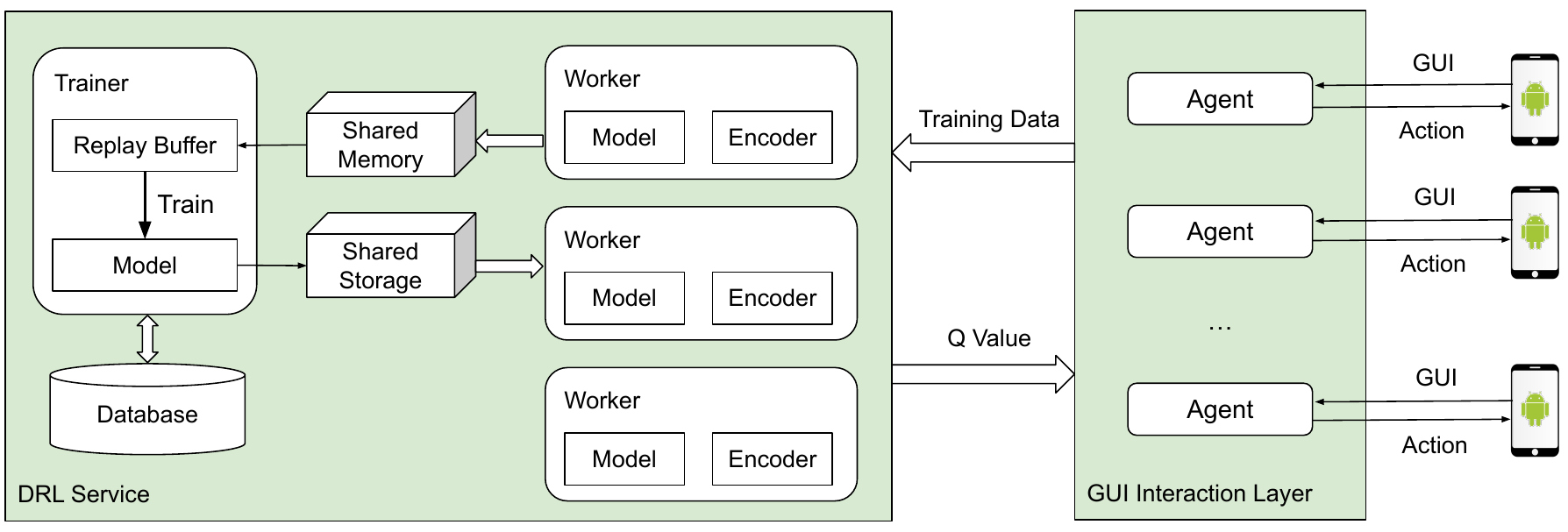}
    \caption{Deep Reinforcement Learning Workflow}
    \label{fig:drl}
\end{figure*}

We train a deep reinforcement learning model to select the GUI event with the most likelihood to reach target changed functions. The overall architecture of the deep RL model is shown in Figure~\ref{fig:drl}. It consists of two modules, \emph{DRL Service} (the server) and \emph{GUI Interaction Layer} (the client) that are are described below. 

\paragraph{DRL Service} The DRL service is responsible for model training and Q value calculation. It comprises the following components:

\begin{itemize}
    \item \textbf{Shared Memory} transfers the training data to the replay buffer of the model trainer.
    \item \textbf{Shared Storage} stores shared variables across processes.
    \item \textbf{Trainer} trains the deep reinforcement learning model. It stores the training data using the replay buffer and continuously samples the training data to train the model. After each training, it saves model parameters to the shared storage and the database.
    \item \textbf{Worker} provides two types of interfaces, \texttt{Add Training Data} and \texttt{Get Q value} to the GUI interaction layer. The model represents the up-to-date trained model fetched periodically from the shared memory and the encoder translates the raw GUI information to vectors that is digestible by the model. When \texttt{Add training data} is called, the worker stores encoded GUI action and functions to the shared memory and when \texttt{Get Q value} is called, the worker gives the encoded GUI information to the model and gets inferred GUI action to be performed.
\end{itemize}

\paragraph{GUI Interaction Layer} The GUI interaction layer is responsible for device interactions and supports multiple device training and testing.
For each available device, the GUI interaction layer instantiates a corresponding agent to communicate with it using the Android Debugging Bridget that provides capabilities including GUI screen query and event execution.
The GUI screen is represented in an XML format where each node stands for a GUI element with its class (type of the widget), resource ID (identifier of the GUI element), text, position, etc.

The agent continuously queries the device screen to get available GUI elements and sends this GUI information to the encoder. When a decision is made by the deep reinforcement learning service, it gets and executes the best GUI event with the highest Q value.

\subsubsection{Model Training}
\label{sec:training}

The training phase adopts a centralized training and distributed execution architecture, with each device corresponding to a Worker and multiple Workers corresponding to a Trainer.
The specific process is as follows:

\begin{enumerate}
    \item[i.] Each device installs the App under test, randomly selects a function as the target function, continuously obtains the current screen's GUI information and the list of functions covered by the previous action, and sends them to the corresponding Worker. The Worker parses the available action list from the current screen and selects the action generation method using the $\epsilon-greedy$ method. With a probability of $\epsilon$, a random action is selected and returned to the client for execution. With a probability of $1-\epsilon$, the local model is called to obtain the Q value of each action, and the action with the highest Q value is returned to the client for execution.
    \item[ii.] As the Worker continuously receives states and selects actions, it stores the test sequence information in the Shared Memory. The stored content is $$<s_0,a_0,F_0,s_1,a_1,F_1,...,s_t,a_t,F_t,s_{t+1}>$$ 
    
    with a maximum sequence length of N, where $s_t$ is the current state's GUI information, including the current screen's activity and XML; $a_t$ is the action executed on the current screen, $F_t$ is the list of functions that can be covered after executing $a_t$ in $s_t$, and $s_{t+1}$ is the next screen after executing $a_t$ in $s_t$.
    \item[iii.] The Trainer monitors and reads the sequences in the Shared Memory in real-time and generates training data stored in the Replay Buffer. The training data takes the form, $$(g_t,s_t,a_t,r_t,s_{t+1},d_t)$$
    
    where $g_t$ is a specific function that is the directed target, $r_t$ is the reward value, and $d_t$ is the termination flag. The training data is generated using the stored sequence by traversing the entire sequence. For each $(s_t,a_t)$ pair in the sequence, the following steps are repeated $K$ times, \\ 
    1. Randomly select a function from the list of all functions covered as the target (after executing the action) represented as 
    $$g_{t,i}\in F_t\cup F_{t+1}\cup...\cup F_N$$
    
    2. Calculate the reward based on the selected target:
    \begin{itemize}
        \item The executed Action triggers the target Function, $r_t=1$, $d_t=1$
        \item The executed Action does not cause a change in the screen, $r_t=-0.001$, $d_t=0$
        \item The target Function can be triggered within n steps after executing the Action, $r_t=0.01 * \gamma^n$, $d_t=0$
        \item Others, $r_t=-0.0001$, $d_t=0$
    \end{itemize}
    where $\gamma$ is the discount factor.
    \item[iv.] The Trainer continuously extracts training data from the Replay Buffer and updates the model parameters
       $$r_t+\gamma(1-d_t) Q'(s_{t+1},arg max_{a}Q(s_{t+1},a,g_t;\theta),g_t;\theta')$$
       $$\rightarrow Q(s_t,a_t,g_t;\theta)$$

    where $Q'$ is the target network, its parameters are $\theta'$, and the parameters are periodically copied from the prediction network.
    \item[v.] During the training process, the model is saved to the cloud and Shared Storage simultaneously. The Worker regularly obtains the latest model from Shared Storage and replaces the local model.
\end{enumerate}

\begin{figure*}[htb]
    \centering
    \includegraphics[scale=0.45]{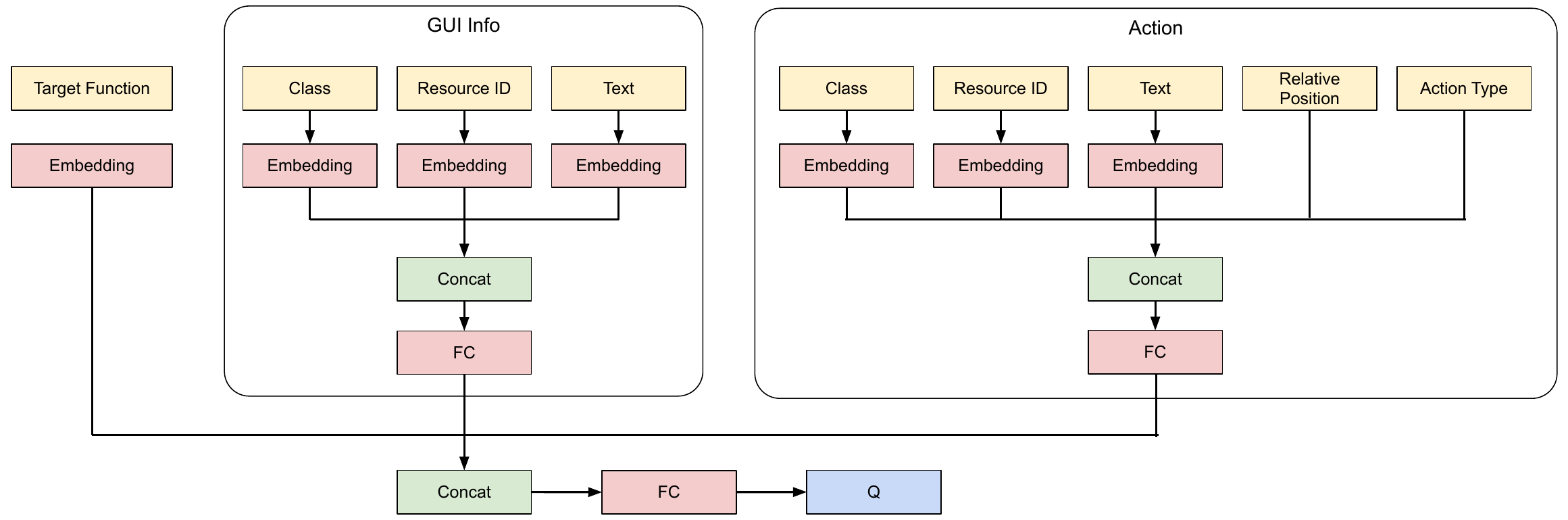}
    \caption{Deep Reinforcement Learning Model Structure}
    \label{fig:model}
\end{figure*}

The deep reinforcement learning model structure is shown in Figure~\ref{fig:model}. 

\subsubsection{Guided GUI Exploration}
\label{sec:exploration}

At the beginning of the GUI exploration, the Worker loads the optimal model from the storage and assigns target functions to each available device. The client sends the target function and the current GUI information to the Worker according to its target function. The Worker parses all the available actions in the current state. The Worker inputs this information into the local model, obtains the Q value for each action, and selects the action with the highest Q value to return to the client for execution. This continues until the target function is covered or the maximum number of directed steps is reached. It then proceeds to the next selected function.

\subsection{Implementation}

\toolname is designed as a comprehensive automated testing framework for Android, comprising code instrumentation, client and server modules.
The code instrumentation tool is developed using Java based on the ASM framework to insert function coverage loggers and send socket messages of real time incremental function coverage to the client.
A Gradle plugin is developed using Groovy to perform code instrumentation automatically in existing Android projects.
The client module is written in Java and leverages the GUI tree retrieving and action execution features of Fastbot2~\cite{lv2022fastbot2} to interact with the App being tested.
The server module, developed in GoLang, facilitates event selection and enables multi-device collaboration (which permits numerous clients to simultaneously test the same App on multiple devices while sharing the same deep reinforcement learning agent). The server provides APIs to receive GUI trees and returns events for execution by the clients.

%% file: experiment.tex
In this section, we evaluate the feasibility and effectiveness of \toolname in generating GUI actions that exercise changed functions.
We use 10 open source Android Apps from the F-Droid App market and 1 commercial Apps from our industry collaborator. A description of the Apps used is provided below,

\noindent \emph{Open source} - We use 10 Android applications from the F-Droid App market~\footnote{\url{https://f-droid.org/}} that has a catalogue of free and open source Android applications. 
We selected top rated Apps in F-Droid with a well documented commit history across multiple versions. Table~\ref{table:apps} lists the names and versions for the Android Apps used in our experiment.
For each open source App, we selected five commits after the base version that resulted in updates on a Java or Kotlin method of the Apps.

\noindent \emph{Commercial} - We use a popular enterprise collaboration software, Lark, developed by our industrial collaborator (\#11 in bold in Table~\ref{table:apps}). The code instrumentation and APK building is done by their developers. 
We compare with state-of-the-art (SOTA) Android GUI testing tools, \textit{ARES}\cite{romdhana2022deep} and a commercial company in-house tool, \textit{Fastbot2}.
\begin{table}[htbp]
\centering
\caption{Subject Apps: 10 open source Apps and 1 commercial App (listed at \#11)}
\label{tab:apps}
\begin{tabular}{|c|l|l|l|}
\hline
\textbf{\#} & \textbf{App Name} & \textbf{Base Version} & \textbf{APK Size} \\ \hline
1 & Amaze File Manager & 3.8.1 & 15.43 MB \\ \hline
2 & Diary & 1.77 & 0.58 MB \\ \hline
3 & OpenTasks & 1.2.3 & 0.30 MB \\ \hline
4 & Simple Draw & 6.8.0 & 5.56 MB \\ \hline
5 & Simple File Manager & 6.14.0 & 3.73 MB \\ \hline
6 & WiFiAnalyzer & 3.0.1 & 3.48 MB \\ \hline
7 & Hibi & 1.3.1 & 11.98 MB \\ \hline
8 & Suntimes & 0.14.12 & 40.12 MB \\ \hline
9 & Nani & 0.3.0 & 8.20 MB \\ \hline
10 & Currency & 1.31 & 4.43 MB \\ \hline
11 & \textbf{Lark} & 5.15.1 & 183.7 MB \\ \hline
\end{tabular}
\label{table:apps}
\end{table}

\noindent We investigate the following research questions: 
\begin{description}[topsep = 0pt, itemsep = 0pt]
\item[Q1. \toolname versus \textit{Fastbot2} exploration:] 
\textit{Is \toolname able to exercise the selected target functions better than \textit{Fastbot2}'s approach that randomly explores functions in the source code?} 
	
	To answer this question, we first randomly selected 50 functions within each open source App as our target functions.
 Then, we used the \toolname and the \textit{Fastbot2} approach to each generate 1000 events to exercise the App Under Test (AUT). After each test, we recorded the number of target functions covered by the events and compared the result between \toolname and \textit{Fastbot2}.
	
	\item[Q2. Efficiency in testing commit associated changes:] 
	\textit{Is \toolname more effective in exercising changed functions associated with commits in open source Apps when compared to state-of-the-art tool \textit{ARES} and baseline method, \textit{Fastbot2}?}

    For each commit, we manually identified changed functions associated with it.  We then ran \toolname, \textit{ARES}, and \textit{Fastbot2}  to generate 1000 events each to exercise the AUT.
    The number of events required by each approach to exercise the changed functions associated with the commit were recorded. The approach that required the lowest number of events to exercise all the changed functions was deemed the most effective for that particular commit. In case any of the tools failed to exercise all the changed functions within 1000 events, the result was recorded as 1000.
		 
	\item[Q3. Testing effectiveness on commercial Apps:] 
	\textit{Is \toolname better than \textit{Fastbot2} in exercising changed functions within the commercial App, Lark?}
	For this question, we use developer identified changed functions for updates in Lark. As the commercial App is large and complex, we set 200 GUI events as the maximum number of attempts for each changed function and report whether these functions are covered by \toolname. We then execute \toolname and \textit{Fastbot2} to assess number of events required to execute all changed functions.

        \item[Q4. When is \toolname effective:]
        \textit{What factors impact the effectiveness of \toolname in exercising changed functions?}
        We conduct a study using the commercial App, Lark, to better understand the reasons for \toolname being effective at exercising some changed functions but not others.  
        We randomly select 100 listener functions (callback functions that are directly bound to GUI events, such as \texttt{onClickListener}) and check whether \toolname is able to exercise them within 200  GUI events for each function. In contrast to the random selection, we also sample functions according to their heat -- defined as the triggered frequency during training.  We report the success rate of exercising heat-based sampled functions within the same 200-events budget as used with the random selection.
\end{description}

\noindent \textbf{Selected Tools.} We select test input generation tool, \textit{ARES} for comparison in research question 2 since it is the SOTA reinforcement learning-based GUI testing tool. We attempted to use \textit{ARES} for comparison over the commercial App in research question 3  but found that the time needed to generate events over the commercial App was immeasurably long and the tool timed out after generating very few events. 
\textit{Fastbot2} is a reusable automated model-based Android GUI testing tool that leverages probabilistic model and reinforcement learning to utilise prior exploration data and prioritise new GUI states that was not covered by previous testing runs. \textit{Fastbot2} is not designed to target changes.  We compare with \textit{Fastbot2} as it is the in-house testing tool used by the Lark developer team. 

As mentioned in Section~\ref{sec:regression}, it is not possible to make a comparison with CAT and ATUA with equivalent testing capabilities for change-focused testing, since the static analysis module they depend on is unable to handle large applications, even when utilising a devbox with 128 GB of RAM.

\textbf{Experiment Platform.} We conducted our experiment on multiple Samsumg Galaxy S10 devices running Android 11 systems. \toolname and \textit{Fastbot2} were run on the mobile devices, while ARES was run on a MacBook Pro equipped with a 2.6GHz Intel Core i7 Processor.

%% file: result.tex
We present results from our experiment in the context of the research questions in Section~\ref{sec:experiment}.
\subsection{HAWKEYE versus Fastbot2 Exploration}

For the first research question, we assess if randomly selected 50 target functions in each of the open source Apps can be exercised better by \toolname than \textit{Fastbot2}. Function coverage over the 50
target functions is reported for both tools over each App in Table~\ref{tab:target-coverage} in the columns. The final column in Table~\ref{tab:target-coverage} shows the number of target functions covered by both \toolname and \textit{Fastbot2}. On average, across the Apps, we find \toolname outperformed Fastbot2 in terms of target function coverage. 
Specifically, \toolname covered an average of 45 out of 50 target functions, whereas Fastbot2 covered only 39. 
Furthermore, \toolname achieved a higher function coverage for six out of the ten open source Apps than \textit{Fastbot2}.
\toolname and \textit{Fastbot2} both achieved a 100\% coverage on the WiFi Analyzer App. This is because the target functions for this App happened to be easily accessible through only a few actions on the MainActivity.

Although \textit{Fastbot2} performed comparably to \toolname in many of the applications, significant performance differences were observed on \texttt{Currency, Hibi, Suntimes} where \toolname exercised 12--22 functions more on each of these Apps. By manual inspection, we found that this is because the target functions of these Apps require more steps on average to exercise. 
The results indicate that \toolname is more effective in generating events that exercise a broader range of target functions in the AUT.


Furthermore, it is worth noting that the intersection column in Table~\ref{tab:target-coverage} indicates a large overlap between the covered functions of \toolname and \textit{Fastbot2}. This observation suggests that \toolname has the capability to exercise additional functions in the AUTs, thereby enabling more comprehensive testing of the applications.

\begin{table}[htbp]
    \centering
    \caption{Target Functions Coverage}
    \label{tab:target-coverage}
    \begin{tabular}{|l|c|c|c|}
        \hline
        \textbf{App Name} & \textbf{Fastbot2} & \textbf{\toolname} & \textbf{Intersection} \\ \hline
        Amaze File Manager & \textbf{37} & 33 & 31 \\ \hline
        Currency & 32 & \textbf{45} & 32 \\ \hline
        Diary & 40 & \textbf{43} & 39 \\ \hline
        Hibi & 17 & \textbf{39} & 16 \\ \hline
        Nani & 49 & \textbf{50} & 49 \\ \hline
        OpenTasks & \textbf{50} & 39 & 39 \\ \hline
        Simple Draw & \textbf{28} & 27 & 22 \\ \hline
        Simple File Manager & 39 & \textbf{49} & 39 \\ \hline
        Suntimes & 25 & \textbf{37} & 22 \\ \hline
        WiFi Analyzer & \textbf{50} & \textbf{50} & 50 \\ \hline
        Avg. & 36.7 & \textbf{41.2} & N/A \\ \hline
    \end{tabular}
\end{table}

\begin{table}[h]
\centering
\begin{tabular}{|c|c|c|c|c|c|}
\hline
\multirow{2}{*}{\begin{tabular}[c]{@{}c@{}}App\\ Index\end{tabular}} & \multirow{2}{*}{\begin{tabular}[c]{@{}c@{}}\#Changed\\Functions\end{tabular}} &\multirow{2}{*}{\textbf{Commit}} & \multirow{2}{*}{\textbf{Fastbot2}} & \multirow{2}{*}{\textbf{\toolname}} & \multirow{2}{*}{\textbf{ARES}} \\&  & &  &  & \\ \hline
\multirow{5}{*}{1}&1&6d6a192 & 1000 & 1000 & 1000 \\ \cline{2-6} 
&1& 92256a7 & 1000 & 54 & \textbf{6} \\ \cline{2-6} 
&6& 27d1e2b & 1000 & 1000 & 1000 \\ \cline{2-6} 
&3& 67d6712 & 1000 & 1000 & 1000 \\ \cline{2-6} 
&33& 538fd8a & 1000 & 1000 & 1000 \\ \hline
\multirow{5}{*}{2}&3&c4aabf2 & 309 & 399 & \textbf{115} \\ \cline{2-6} 
&1& 54c1335 & \textbf{354} & 487 & 873 \\ \cline{2-6} 
&1& 7c51891 & 55 & \textbf{4} & 21 \\ \cline{2-6} 
&1& 2070eb0 & 305 & \textbf{5} & 6 \\ \cline{2-6} 
&1& f66a966 & \textbf{71} & 81 & 197 \\ \hline
\multirow{5}{*}{3}&1&f53cddd & 1000 & 1000 & \textbf{633} \\ \cline{2-6} 
&1& 1a1669f & 1000 & 1000 & \textbf{109} \\ \cline{2-6} 
&1& f175a71 & 1000 & \textbf{773} & 1000 \\ \cline{2-6} 
&2& 700a773 & 1000 & 701 & \textbf{20} \\ \cline{2-6} 
&4& 47d6676 & 1000 & 1000 & 1000 \\ \hline
\multirow{5}{*}{4}&2&fd7bca5 & 630 & \textbf{9} & 754 \\ \cline{2-6} 
&1& 621d932 & 1000 & 1000 & 1000 \\ \cline{2-6} 
&6& c660f5d & 1000 & 1000 & 1000 \\ \cline{2-6} 
&6& 4e2e9f7 & 1000 & 1000 & 1000 \\ \cline{2-6} 
&3& c1c9be2 & 1000 & 1000 & 1000 \\ \hline
\multirow{5}{*}{5}&2&ea56927 & 1000 & 1000 & 1000 \\ \cline{2-6} 
&4& 440df4e & 1000 & 1000 & 1000 \\ \cline{2-6} 
&3& 0488984 & 40 & 27 & \textbf{3} \\ \cline{2-6} 
&2& ce18931 & 1000 & 1000 & 1000 \\ \cline{2-6} 
&1& 0b5d6ae & 1000 & 1000 & \textbf{183} \\ \hline
\multirow{5}{*}{6}&22&b8c4544 & 1000 & 1000 & 1000 \\ \cline{2-6} 
&1& e750aca & \textbf{3} & 10 & 5 \\ \cline{2-6} 
&51& 8ddf52d & 1000 & 1000 & 1000 \\ \cline{2-6} 
&1& d7f08ff & 1000 & 1000 & 1000 \\ \cline{2-6} 
&5& 238fae0 & 1000 & 1000 & 1000 \\ \hline
\multirow{5}{*}{7}&53&2c4dccd & 1000 & 1000 & 1000 \\ \cline{2-6} 
&2& 294609c & 1000 & 1000 & 1000 \\ \cline{2-6} 
&1& 944848a & 1000 & 1000 & 1000 \\ \cline{2-6} 
&2& b7f113b & 240 & \textbf{4} & 25 \\ \cline{2-6} 
&4& 1456648 & \textbf{850} & 1000 & 1000 \\ \hline
\multirow{5}{*}{8}&1&69ccd65 & \textbf{462} & 524 & 1000 \\ \cline{2-6} 
&4& 4dfa28c & 1000 & 1000 & 1000 \\ \cline{2-6} 
&11& 69db31a & 1000 & 1000 & 1000 \\ \cline{2-6} 
&2& 2f0e4a3 & 165 & 133 & \textbf{6} \\ \cline{2-6} 
&2& ab47bca & 1000 & 1000 & 1000 \\ \hline
\multirow{5}{*}{9}&3&f833032 & 1000 & 1000 & 1000 \\ \cline{2-6} 
&1& 95d6d30 & \textbf{2} & 3 & 563 \\ \cline{2-6} 
&5& 3b27600 & 1000 & 1000 & 1000 \\ \cline{2-6} 
&27& ad10290 & 1000 & 1000 & 1000 \\ \cline{2-6} 
&12& cbdf999 & 1000 & 1000 & 1000 \\ \hline
\multirow{5}{*}{10}&30&3ba6c7c & 1000 & 1000 & 1000 \\ \cline{2-6} 
&5& 7156da1 & 488 & \textbf{20} & 1000 \\ \cline{2-6} 
&1& 966b094 & 52 & 196 & \textbf{40} \\ \cline{2-6} 
&1& 81e79d8 & 382 & 1000 & \textbf{277} \\ \cline{2-6} 
&3& dc3bf39 & 22 & 3 & \textbf{2} \\ \hline
\multirow{2}{*}{Avg.} & \multirow{2}{*}{6.8} & & \multirow{2}{*}{261} & \multirow{2}{*}{\textbf{191}} & \multirow{2}{*}{202} \\ &&&&&\\ \hline
\multirow{2}{*}{\begin{tabular}[c]{@{}c@{}}Success\\ Rate\end{tabular}}  & & & \multirow{2}{*}{34\%} & \multirow{2}{*}{36\%} & \multirow{2}{*}{\textbf{38\%}} \\ &&&&&\\\hline
\end{tabular}
    \caption{Changed Function Coverage on Open-source Apps.}
    \label{tab:trigger-rate}
\end{table}

\subsection{Efficiency in Testing Commit-associated Changes}
Table~\ref{tab:trigger-rate} shows commits for each open source App by \toolname, \textit{Fastbot2} and \textit{ARES}. We use five recent subsequent commits of base version with code changes submitted for each App. The second column shows the number of changed functions for each commit. There are 6.8 functions updated by a commit on average. We recorded the number of GUI events required for each approach to exercise the changed functions associated with the commit. We used 1000 events as the stopping criterion for the tools. The cells in Table~\ref{tab:trigger-rate} that have 1000 indicate the corresponding tool failed to exercise the changed functions for that commit. Numbers less than 1000 are considered a success for the tool, with lower number of events indicating the tool is more efficient in reaching and exercising the changes. On each commit, the winning tool is emphasized in bold as the one with the lowest number of events. 

It is clear from Table~\ref{tab:trigger-rate}, that all three tools fail to exercise changes on a majority of the commits examined.  Our manual analysis revealed that this is largely due to a significant portion (18\%) of the updated methods being callable only under specific Android or database versions, making them inaccessible when the Android or database version of the AUT does not match.
Also, in four of the \texttt{Suntimes}'s commits, exercising the target functions requires adding App widget on the home screen, which is very unlikely to happen because all the tools tends to relaunch the App when the it accidentally exits the App screen.
\texttt{Currency} is the only App where all 5 commits were successfully exercised by all three tools. This is because the number of updated functions associated with the commits for \texttt{Currency} are small and not version specific, i.e., the functions are not only accessible on devices with specific Android or database versions. Furthermore, these functions require a small number of steps to exercise. 

Success rates for the tools (last row of Table~\ref{tab:trigger-rate}), measured as the fraction of commits that needed less than 1000 events, are comparable. In terms of the average number of events needed for each tool to successfully exercise the commits (second to last row of Table~\ref{tab:trigger-rate}),
\toolname required the fewest events on average -- 191 events -- to exercise the updated methods of a commit, while \textit{ARES} required 202 events and \textit{Fastbot2} required 261 events.

\subsection{Testing Effectiveness on Commercial App}

Table~\ref{tab:rq3} shows twelve commits for the Lark App, the number of changed functions identified by the developers for each commit, and the function coverage achieved by \toolname and the baseline \textit{Fastbot2}. The commits were selected from merge requests that happened during a one week time period and provided by Lark developers. Across the 12 commits, \textit{Fastbot2} performs poorly, achieving less than 10\% coverage on 11 out of the 12 commits. \toolname performs considerably better than \textit{fastbot2} on 8 of the 12 commits with improvements in the range of $1.3 - 90.9\%$. \textit{Fastbot2} performs better than \toolname on 2 out of the 12 commits. 
Function coverage with \toolname is 5\% or less for half the commits. This is primarily because of the stopping condition we imposed of 200 events per changed function. We imposed this condition so we obtain results within a reasonable time. For a commercial App the size of Lark, 200 events is quite small and thus the low coverage observed.  Despite the low function coverage observed with this stopping condition on Lark, \toolname stands out as the most promising tool that can scale to the complexity of a commercial App while being partially successful in exercising updates.  We are confident that the coverage will improve if this limit is lifted. We acknowledge this is a first step in testing commit associated changed functions for commercial Apps with more advances needed in tuning the reinforcement learning technique to better exercise the changes. Nevertheless, \toolname's performance is considerably better than the in-house company tool, \textit{Fastbot2} and it was very well received by the developers who found it saved them time and effectively exercised some changes.

\begin{table}[]
\centering
\begin{tabular}{|c|c|cc|cc|}
\hline
\multirow{2}{*}{\begin{tabular}[c]{@{}c@{}}Commit\\ ID\end{tabular}} & \multirow{2}{*}{\begin{tabular}[c]{@{}c@{}}\# Changed\\ Functions\end{tabular}} & \multicolumn{2}{c|}{Fastbot2}                                                                                                       & \multicolumn{2}{c|}{HawkEye}                                                                                                        \\ \cline{3-6} 
                                                                     &                                                                                 & \multicolumn{1}{c|}{\begin{tabular}[c]{@{}c@{}}\#\\ Covered\end{tabular}} & \begin{tabular}[c]{@{}c@{}}Coverage\\ (\%)\end{tabular} & \multicolumn{1}{c|}{\begin{tabular}[c]{@{}c@{}}\#\\ Covered\end{tabular}} & \begin{tabular}[c]{@{}c@{}}Coverage\\ (\%)\end{tabular} \\ \hline
1                                                                    & 11                                                                              & \multicolumn{1}{c|}{0}                                                    & 0.0\%                                                   & \multicolumn{1}{c|}{\textbf{10}}                                          & \textbf{90.9\%}                                         \\ \hline
2                                                                    & 8                                                                               & \multicolumn{1}{c|}{\textbf{4}}                                           & 50.0\%                                                  & \multicolumn{1}{c|}{2}                                                    & 25.0\%                                                  \\ \hline
3                                                                    & 68                                                                              & \multicolumn{1}{c|}{0}                                                    & 0.0\%                                                   & \multicolumn{1}{c|}{\textbf{34}}                                          & \textbf{50.0\%}                                         \\ \hline
4                                                                    & 14                                                                              & \multicolumn{1}{c|}{\textbf{1}}                                           & \textbf{7.1\%}                                          & \multicolumn{1}{c|}{0}                                                    & 0.0\%                                                   \\ \hline
5                                                                    & 393                                                                             & \multicolumn{1}{c|}{0}                                                    & 0.0\%                                                   & \multicolumn{1}{c|}{\textbf{5}}                                           & \textbf{1.3\%}                                          \\ \hline
6                                                                    & 138                                                                             & \multicolumn{1}{c|}{0}                                                    & 0.0\%                                                   & \multicolumn{1}{c|}{\textbf{17}}                                          & \textbf{12.3\%}                                         \\ \hline
7                                                                    & 209                                                                             & \multicolumn{1}{c|}{0}                                                    & 0.0\%                                                   & \multicolumn{1}{c|}{\textbf{15}}                                          & \textbf{7.2\%}                                          \\ \hline
8                                                                    & 216                                                                             & \multicolumn{1}{c|}{0}                                                    & 0.0\%                                                   & \multicolumn{1}{c|}{0}                                                    & 0.0\%                                                   \\ \hline
9                                                                    & 36                                                                              & \multicolumn{1}{c|}{0}                                                    & 0.0\%                                                   & \multicolumn{1}{c|}{0}                                                    & 0.0\%                                                   \\ \hline
10                                                                   & 59                                                                              & \multicolumn{1}{c|}{0}                                                    & 0.0\%                                                   & \multicolumn{1}{c|}{\textbf{10}}                                          & \textbf{16.9\%}                                         \\ \hline
11                                                                   & 40                                                                              & \multicolumn{1}{c|}{0}                                                    & 0.0\%                                                   & \multicolumn{1}{c|}{\textbf{14}}                                          & \textbf{35.0\%}                                         \\ \hline
12                                                                   & 133                                                                             & \multicolumn{1}{c|}{2}                                                    & 1.5\%                                                   & \multicolumn{1}{c|}{\textbf{7}}                                           & \textbf{5.3\%}                                          \\ \hline
\end{tabular}
    \caption{Changed Function Coverage on Lark}
    \label{tab:rq3}
\end{table}

\begin{table*}[!h]
\centering
\begin{tabular}{|l|l|l|l|l|l|l|}
\hline
\textbf{Function Heat Range}            & \textgreater{}=300 & 100-300        & 50-100         & 20-50 & 10-20 & \textless{}10 \\ \hline
\textbf{Number of Functions}            & 196                & 192            & 159            & 260   & 208   & 558           \\ \hline
\textbf{Success Rate \textgreater 30\%} & 76.5\% (150/196)    & 29.2\% (56/192) & 11.9\% (19/159) & -     & -     & -             \\ \hline
\end{tabular}
    \caption{Case Study on Lark}
    \label{tab:rq4}
\end{table*}

\subsection{Case Study of the Effectiveness of \toolname}

\subsubsection{Listener Function Coverage}

We randomly selected 100 listener functions from Lark as target functions and set the maximum number of GUI events for each function as 200 and repeat this experiment 13 times.

\toolname is able to cover, on average, 52 out of 100 functions, with an average of 42  GUI events. Across all the 13 testing runs, \toolname covers 89 out the 100 functions.
It is worth noting that, \toolname is able to cover 44 functions very quickly within the first 5 minutes. However, \toolname exhausted the entire 3-hour budget to cover the rest of the functions.
As with research question 3, we believe that function coverage will improve if \toolname uses more than 200 GUI events for each function given the complexity and size of Lark. 

\subsubsection{Function Coverage based on Heat}

As shown in Table~\ref{tab:rq4}, after 10 hours of model training based on random exploration, 196 functions are exercised more than 300 times, 192 functions are exercised between 100 and 300 times, 159 functions are exercised between 50 and 100 times, 260 functions are exercised between 20 and 50 times, 208 functions are exercised between 10 and 20 times and 558 functions are exercised less than 10 times.

As might be expected, We find that the function coverage achieved by \toolname is correlated to the function heat --- more frequently explored (more than 300 times) functions by the training model are covered more easily by \toolname's tests. 
On the other hand, \toolname struggles to cover functions that were seen by the training model less than 50 times. The correlation with exploration frequency is not unexpected and we will aim to improve model training by visiting a larger set of functions more frequently in the future. Our current training was limited by resource availability and we believe this issue could be resolved by increasing the number of devices used for training to cover more GUI states and functions.

\section{Threats to Validity}
A potential threat to internal validity is bugs in \toolname's implementation. To mitigate this threat, we conducted careful code reviews and extensive testing. When used on the commercial App, the developers at the commercial site conducted several rounds of manual inspection to mitigate the risk of manual mistakes or omissions in the tool. 


A potential threat to the external validity is related to the fact that
the set of Android Apps we have considered in this study may not
be an accurate representation of a potential App under test. We attempt to reduce the selection
bias by using a dataset of 10 open source Apps from different categories with a variety of Android features, and one large, complex commercial App.

A threat to construct validity is caused by restricting the number of GUI events generated by the tools -- \toolname, \textit{Fastbot2, ARES} -- to 1000 events for open source Apps and 200 for the commercial App. Restriction to 1000 input events is inspired from related work in GUI testing~\cite{li2017droidbot, borges2018droidmate} that used this in their default settings. 

A final threat to validity is the limited number of tools used in comparison.
As discussed in Section~\ref{sec:regression}, we cannot compare with CAT and ATUA with similar capabilities to test changes as the static analysis module they rely on fail on large Apps even using a devbox with 128 GB RAM.
We used \textit{Fastbot2} and \textit{ARES} for comparison  as they are the SOTA RL-based GUI testing tools that we are aware of. \textit{Fastbot2} is an industry developed GUI testing tool unlike \textit{ARES}, so the comparison tools do have some diversity in that regard. 

%% file: conclusion.tex
In this paper, we introduced \toolname, specifically designed for generating GUI test inputs aimed at changed functions from Android App updates.
\toolname uses deep reinforcement learning to learn the mapping between GUI events and functions to produce GUI event sequences to interact with these targeted changed functions.
We conducted an empirical evaluation of the performance of \toolname by comparing it with \textit{ARES} and \textit{Fastbot2} across 10 open-source Android Apps and 1 large commercial App. Our evaluation yielded the following observations.

\begin{enumerate}
    \item \toolname can execute changed functions more frequently than RL testing tools, \textit{Fastbot2} and \textit{ARES}, while using fewer GUI events.
    \item For a complex commercial App, Lark, \toolname exhibits better performance than \textit{Fastbot2} in covering listener functions. \toolname covers 85\% of Lark's randomly selected listener functions within the first 180 minutes of a  testing run. 
    \item \toolname is effective at exercising functions that were commonly observed during model training.
\end{enumerate}

\vfill\null